\documentclass[conference]{IEEEtran}
\IEEEoverridecommandlockouts
\usepackage{cite}
\usepackage{amsmath,amssymb,amsfonts}
\usepackage{algorithmic}
\usepackage{graphicx}
\usepackage{textcomp}
\usepackage{xcolor}
\usepackage{enumerate}
\usepackage{array,booktabs,arydshln,xcolor}
\usepackage[hidelinks]{hyperref}
\usepackage{caption}
\usepackage{comment}
\usepackage{siunitx}
\usepackage{stfloats}

\newcommand{\cmmnt}[1]{}

\def\BibTeX{{\rm B\kern-.05em{\sc i\kern-.025em b}\kern-.08em
    T\kern-.1667em\lower.7ex\hbox{E}\kern-.125emX}}
    
\begin{document}

\title{A Tsetlin Machine-driven Intrusion Detection System for Next-Generation IoMT Security \\
}

\author{
    \IEEEauthorblockN{Rahul Jaiswal, Per-Arne Andersen, Linga Reddy Cenkeramaddi, Lei Jiao~and Ole-Christoffer Granmo\\
    The Centre for Artificial Intelligence Research (CAIR) \\
    \IEEEauthorblockA{Department of ICT, University of Agder, Norway}
{\{rahul.jaiswal, per.andersen, linga.cenkeramaddi, lei.jiao, ole.granmo\}@uia.no}
}}

\maketitle

\begin{abstract} 
The rapid adoption of the Internet of Medical Things (IoMT) is transforming healthcare by enabling seamless connectivity among medical devices, systems, and services. However, it also introduces serious cybersecurity and patient safety concerns as attackers increasingly exploit new methods and emerging vulnerabilities to infiltrate IoMT networks. This paper proposes a novel Tsetlin Machine (TM)-based Intrusion Detection System (IDS) for detecting a wide range of cyberattacks targeting IoMT networks. The TM is a rule-based and interpretable machine learning (ML) approach that models attack patterns using propositional logic. Extensive experiments conducted on the CICIoMT-2024 dataset, which includes multiple IoMT protocols and cyberattack types, demonstrate that the proposed TM-based IDS outperforms traditional ML classifiers. The proposed model achieves an accuracy of 99.5\% in binary classification and 90.7\% in multi-class classification, surpassing existing state-of-the-art approaches. Moreover, to enhance model trust and interpretability, the proposed TM-based model presents class-wise vote scores and clause activation heatmaps, providing clear insights into the most influential clauses and the dominant class contributing to the final model decision.
\end{abstract}


\section{Introduction} 
\label{intro}

\begin{figure}[t!]
\centering
\includegraphics[width=\columnwidth,height=5.5cm,keepaspectratio]{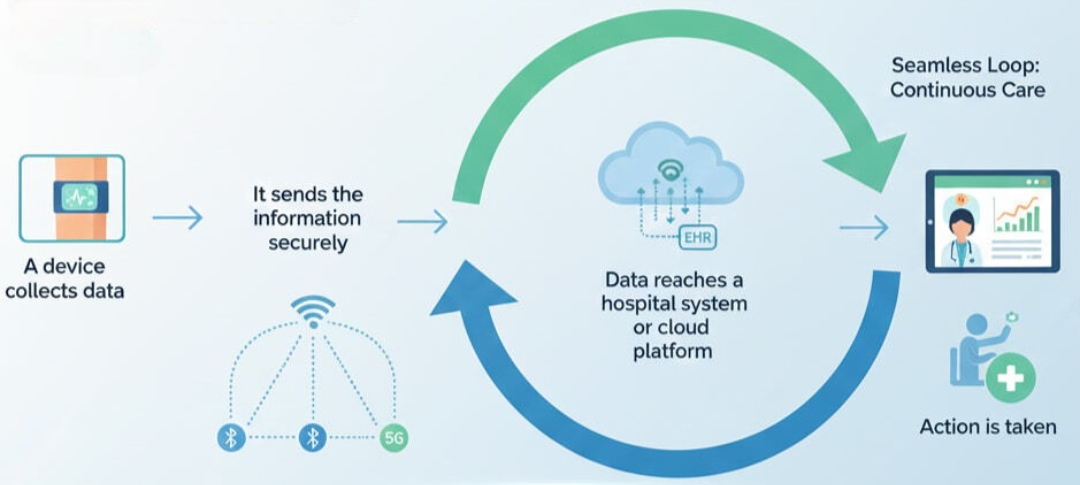} 
\caption{A simple illustration of IoMT working in healthcare.}
\label{iomt_app} \vspace{-3mm}
\end{figure}

The COVID-19 pandemic triggered a dramatic increase in online doctor consultations as hospitals faced high patient loads and capacity constraints. Consequently, patient health data began to be transmitted to hospitals and physicians through digital mediums, such as wearable medical devices, as shown in Fig.~\ref{iomt_app}. The ecosystem connecting medical devices, healthcare applications, and digital systems to communicate over the Internet is referred to as the Internet of Medical Things~(IoMT)~\cite{razdan2022internet}. It ensures that crucial medical data flows seamlessly from the patient device to the physician, enhancing both the speed and the quality of continuous patient care. 

A recent report on the IoMT global business market~\cite{report_iomt} states that the United States alone accounted for an IoMT market size of USD 230.69 billion in 2024 across software, hardware, and services, and is projected to grow at a compound annual growth rate (CAGR) of 18.2\% from 2025 to 2030, reaching USD 658.57 billion. In a similar trend, the European and Asian IoMT markets are expected to grow at CAGRs exceeding 16\% and 21\%, respectively, during 2025-2030. This indicates that the IoMT growth underscores a major transformation in the healthcare sector, fueled by the rising use of remote patient monitoring, wearable medical devices (e.g., smart watches, fitness trackers), and telehealth services.

IoMT devices are connected to the Internet to send highly sensitive and private medical data. Therefore, they are vulnerable to cyberattacks and introduce security challenges related to data integrity, availability, safety, and patient privacy. For example, manipulated thermometer readings introduced by cyberattackers into hospital systems can compromise patient safety by leading physicians to make incorrect and potentially life-threatening medication decisions. The CrowdStrike Global Threat Report~\cite{report_threat} states that healthcare systems accounted for 9\% of intrusion attacks worldwide in 2025.

To mitigate cyberattacks on IoMT environments, Intrusion Detection Systems (IDS) are widely deployed~\cite{iwendi2021security}. These systems employ network-based security mechanisms for monitoring traffic, detecting anomalies/attacks, and identifying potential security breaches in real time. The robust IDS solutions are essential to ensure the confidentiality, integrity, security, and availability of IoMT systems, thereby enabling secure, reliable, and seamless delivery of healthcare services.

This paper proposes a novel IDS system designed to identify diverse network-based attacks on IoMT devices using an explainable machine learning (ML) approach based on the Tsetlin Machine\footnote{The Tsetlin Machine is described in detail under Section~\ref{tmc}.} (TM)~\cite{kundu360, granmo2018tsetlin}. To this end, network-based features are extracted from inbound and outbound IoMT traffic. These features are subsequently fed to the TM model to classify network traffic as \texttt{benign (normal)} or \texttt{malicious attacks}. To enhance trust in the TM model’s decisions and ensure transparency beyond black-box ML classifiers, the interpretability of the TM model is presented.

The remainder of this paper is organized as follows. Section~\ref{rel_work} presents the related work and motivation. Section~\ref{bg} provides a brief overview of the classifiers used in this study. Section~\ref{prop_ids} describes the proposed TM-based intrusion detection system. Section~\ref{exp_data} details the experimental dataset. Section~\ref{res_dis} compares and discusses the results. Section~\ref{con_fut} concludes the paper with future research directions.

\section{Related Work and Motivation} 
\label{rel_work}
IDS systems are essential for protecting sensitive medical data in IoMT environments against cyberattacks. Several studies have been conducted to address these emerging security challenges. Traditionally, rule-based and signature-based IDS~\cite{mitchell2014behavior, nawaal2024signature} methods are used, which rely heavily on static rules and known signatures. However, the dynamic nature of IoMT environments limits their ability to detect new and evolving threats, leaving IoMT devices vulnerable to emerging attack types. Moreover, traditional IDS struggles to handle the large volume of network traffic data generated by IoMT devices, leading to delayed threat response time and increased security vulnerabilities. 

Consequently, machine learning and deep learning-based intrusion detection approaches have emerged to enhance IoMT security. For example, the study in~\cite{anitha2023artificial} evaluates multiple ML methods to detect cyberattacks on the IEEE DataPort dataset, where K-nearest Neighbours (KNN) attains an accuracy of 89.89\% in a binary classification setting. A deep autoencoder-based IDS is proposed for securing IoMT systems using the NF-ToN-IoT dataset in~\cite{awotunde2021deep}, achieving an accuracy of 89\%  in a 10-class classification task. The authors in~\cite{kavkas2025enhancing} proposed deep neural network (DNN) and long short-term memory (LSTM)-based approaches for cyberattack detection in IoMT systems using the CICIoMT24~\cite{dadkhah2024ciciomt2024} dataset. Both models achieve 99\% accuracy for binary classification, while DNN and LSTM attain accuracies of 78\% and 79\%, respectively, for the six-class classification task. The CICIoMT24 developer uses different ML techniques to detect cyberattacks and achieve 99\% accuracy for binary classification and 73.5\% for the six-class classification~\cite{dadkhah2024ciciomt2024}. However, these ML models do not provide insights into how their predictions are derived, raising significant concerns about transparency and interpretability. 

Recently, the TM~\cite{kundu360, granmo2018tsetlin} has emerged as a promising ML approach for IoMT intrusion detection due to its rule-based and interpretable learning framework, which models attack patterns using propositional logic. Moreover, compared to general Internet of Things (IoT), IoMT systems are safety-critical and resource-constrained; therefore, the TM’s lightweight, interpretable, and efficient design makes it well-suited for reliable on-device intrusion detection. For example, the authors in~\cite{abeyrathna2020intrusion, gunvaldsen2023towards} propose anomaly detection frameworks based on the TM and evaluate them on various datasets, reporting higher accuracies than traditional ML classifiers. Motivated by these findings, we propose a novel, transparent, and interpretable TM-based IDS for IoMT environments to identify diverse cyberattacks and enhance patient safety. The main contributions of this paper are summarized as follows:
\begin{itemize}
    \item Design of an effective data-driven TM-based IDS for detecting cyberattacks in IoMT environments. 
    \item Comprehensive evaluation and performance analysis of the proposed model under both binary and multi-class attack detection settings using the CICIoMT24 dataset.
    \item Explainability analysis of the proposed model, illustrating how logical clauses and interpretable decision rules contribute to accurate intrusion detection.
    \item Comparative evaluation with existing studies in the literature. Numerically, we show that the proposed model outperforms the state-of-the-art ML methods. 
    \item Demonstrating the practical viability of TM-based solutions for securing IoMT environments/devices. 
\end{itemize}

\section{Classifier Background} 
\label{bg}
This section provides a brief overview of the various classifiers employed in our experiment.

\subsection{Tsetlin Machine}
\label{tmc}
TM is a new ML technique based on the Tsetlin Automaton (TA)~\cite{granmo2018tsetlin}. It is a rule-based and interpretable ML model that learns logical clauses using propositional logic, making it well-suited for IoMT intrusion detection. It represents cyberattack patterns as human-readable logical expressions, enabling transparent detection of network attacks. The binary feature representation technique of TM makes it highly effective for resource-constrained IoMT devices. Moreover, TM shows high potential with class imbalance data, which is a common characteristic of IoMT intrusion-detection datasets, e.g., CICIoMT24 (see Table~\ref{summary_attack}). Next, we briefly present the mathematical formulation of TM.

Let the IoMT network traffic sample be represented by a binary feature vector after pre-processing and binarization as:
\begin{equation}
\mathbf{x} = (x_1, x_2, \ldots, x_d),    \quad x_i \in \{0,1\}.
\end{equation}

For a multi-class classification with $C$ classes, TM associates each class $c \in \{1,2,\ldots,C\}$ with a set of $m$ clauses. These clauses are divided equally into positive and negative polarity clauses. With $I^{(c)}_j$ and $\bar{I}^{(c)}_j$ denoting the index sets of included and negated literals, respectively, the output of the $j$-th
clause for class $c$ is defined as~\cite{granmo2018tsetlin}:
\begin{equation}
C^{(c)}_j(\mathbf{x}) =
\bigwedge_{i \in I^{(c)}_j} x_i \;\wedge\;
\bigwedge_{k \in \bar{I}^{(c)}_j} \neg x_k.
\end{equation}

For a class $c$, TM calculates the class score by aggregating the clause outputs as:
\begin{equation}
f_c(\mathbf{x}) =
\sum_{j=1}^{m/2} C^{(c)+}_j(\mathbf{x})
-
\sum_{j=1}^{m/2} C^{(c)-}_j(\mathbf{x}),
\label{eq:class_vote}
\end{equation}
where $C^{(c)+}_j$ and $C^{(c)-}_j$ are the positive and negative clauses for class $c$, respectively.

Finally, the predicted class label is obtained as:
\begin{equation}
\hat{y} = \arg\max_{c} f_c(\mathbf{x}).
\end{equation}

Note that each literal within a clause is governed by a TA that learns to either include or exclude the literal. Moreover, the learning process of TM is regulated by the voting threshold parameter $T$, which limits the number of clauses that learn a certain sub-pattern, and the specificity parameter $s$, which controls clause granularity~\cite{granmo2018tsetlin}.

\subsection{Machine Learning Classifiers}
\label{mlc}
Eight different ML classifiers are used in our experiment.

\subsubsection{Decision Tree (DT)}
\label{dtc}
Decision nodes and leaf nodes combine to construct a  DT~\cite{jaiswal2023caqoe}. The decision node comprises multiple branches to handle the outcome of the test sample. The leaf node represents a class, i.e., the result of a decision. DT works on the principle of divide and conquer~\cite{breiman2017classification}.
\subsubsection{Random Forest (RF)}
\label{rfc}
It is an ensemble of multiple decision trees~\cite{jaiswal2023caqoe}. Each tree provides the classification result, and then the forest selects the class having the highest votes~\cite{jedari2015wi}.
\subsubsection{XGBoost}
\label{xgbc}
It is a combination of classification and regression trees~\cite{chen2016xgboost}. It uses a gradient boosting algorithm to optimize the trees by correcting the previous errors. 
\subsubsection{LightGBM (LGBM)}
\label{lgbc}
It is also a gradient-boosting algorithm, which works on leaf-wise tree growth and histogram-based learning for faster training~\cite{ke2017lightgbm}. 
\subsubsection{K-nearest Neighbours (KNN)}
\label{knnc}
It finds the class of a test sample based on the majority votes of the close neighbours in the feature space using Euclidean distance~\cite{jedari2015wi}.
\subsubsection{Naive Bayes (NB)}
\label{nbc}
It is a probabilistic classifier working on Bayes' theorem~\cite{alpaydin2020introduction}. It ignores possible dependencies, such as correlations among the inputs, and reduces a multivariate problem to a set of univariate problems~\cite{jaiswal2023caqoe}.
\subsubsection{Logistic Regression (LR)}
\label{lrc}
It is a linear classifier that uses the logistic function to model the class probability between $0$ and $1$ for the classification problem~\cite{jaiswal2023caqoe}. 
\subsubsection{Neural Network (NN)}
\label{nnc}
It comprises interconnected layers of neurons, which can learn non-linear and complex feature patterns of malicious cyberattacks~\cite{jaiswal2023caqoe}.

\section{Proposed Intrusion Detection System} 
\label{prop_ids}
\begin{figure}[t!]
\centering
\includegraphics[width=0.9\columnwidth,height=5.25cm]{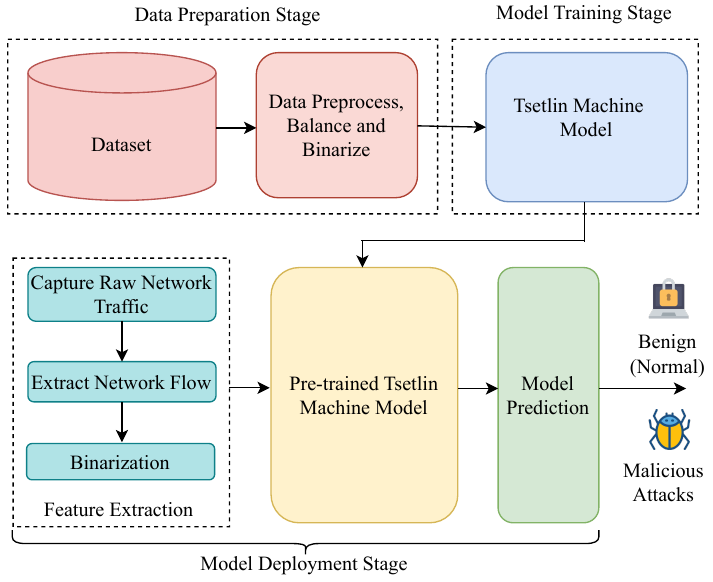} 
\caption{The overall architecture of the proposed TM-based IDS.}
\label{ids_tm} \vspace{-5mm}
\end{figure}

The primary goal of an IDS is to accurately detect and classify diverse cyberattacks in IoMT environments. To achieve this goal, the system design must integrate multiple cooperative components and adopt adaptive and flexible strategies to maximize detection accuracy. The overall architecture of the proposed TM-based IDS is shown in Fig.~\ref{ids_tm}. It mainly comprises three stages. In the first stage, a dataset is selected, and data preparation is performed to preprocess network traffic for training the TM model. It includes data cleaning, handling missing values, class balancing, and binary feature conversion. In the second stage, the TM model is trained, and the learned model is stored for use during deployment. During training, the TM learns interpretable logical patterns in the form of positive and negative clauses using propositional logic and computes class-wise scores for decision-making. To support explainability, the model exposes the activated clauses, providing human-readable rules that contribute to each decision. In the third stage, which is the deployment stage, raw network traffic is captured to derive network flow information. After that, features are extracted, binarized, and input to the pre-trained TM model to classify the network traffic as either \texttt{benign (normal)} or \texttt{malicious attacks}. This prediction can then be forwarded to a firewall to block the traffic originating from malicious sources.

\section{Experimental Dataset} 
\label{exp_data}
To detect different types of cyberattacks, we use an open-source IoMT dataset, CICIoMT24\footnote{Dataset link: https://www.unb.ca/cic/datasets/iomt-dataset-2024.html.}~\cite{dadkhah2024ciciomt2024}, developed by the Canadian Institute for Cybersecurity. It uses different devices and protocols tailored to healthcare security applications for capturing different types of malicious attacks. The dataset is collected from 40 IoMT devices (25 real and 15 simulated devices). Some of the IoMT devices used for developing the dataset are: baby monitor, sleep ring, heart rate arm band, O2 ring, and chest heart rate monitor. A full list of used IoMT devices can be seen in~\cite{dadkhah2024ciciomt2024}. It uses popular IoMT protocols, such as Bluetooth, Message Queuing Telemetry Transport (MQTT), and Wi-Fi. The Bluetooth protocol contains benign (normal) data and a denial-of-service (DoS) attack. The MQTT and Wi-Fi protocols jointly contain benign data and five types of cyberattacks: DoS, distributed denial-of-service (DDoS), reconnaissance (Recon), MQTT, and spoofing. 

\begin{table*} [t!]
\caption{List of features present in CICIoMT24.}
\label{summary_feature} 
\centering
{
\setlength\tabcolsep{1.0pt}
\begin{tabular}{|c|c|c|c|c|c|c|c|}
\hline
\multicolumn{8}{|c|}{Protocol: Bluetooth} \\
\hline
S.No. & Feature name & S.No. & Feature name & S.No. & Feature name & S.No. & Feature name \\
\hline
1 & \texttt{Header\_Length} & 2 & \texttt{Protocol\_Type} & 3 & \texttt{Packet\_Type} & 4 & \texttt{Rate} \\ 
5 & \texttt{HCI\_Command} & 6 & \texttt{HCI\_Event} & 7 & \texttt{HCI\_ACL\_Data} & 8 & \texttt{HCI\_SCO\_Data} \\ 
9 & \texttt{Command\_Complete} & 10 & \texttt{Command\_Status} & 11 & \texttt{LE\_Meta} & 12 & \texttt{Connection\_Complete} \\ 
13 & \texttt{Disconnection\_Complete} & 14 & \texttt{Inquiry\_Complete} & 15 & \texttt{Advertising\_Report} & 16 & \texttt{Read\_Remote\_Features} \\ 
17 & \texttt{Encryption\_Change} & 18 & \texttt{Number\_Completed\_Packets} & 19 & \texttt{Tot\_sum} & 20 & \texttt{Min} \\
21 & \texttt{Max} & 22 & \texttt{AVG} & 23 & \texttt{Std} & 24 & \texttt{Tot\_size} \\ 
\cline{7-8}
25 & \texttt{IAT} & 26 & \texttt{Number} & 27 & \texttt{Variance} & \multicolumn{2}{c|}{Total number of features = 27} \\
\hline
\multicolumn{8}{|c|}{Protocol: MQTT and Wi-Fi} \\
\hline
1 & \texttt{Header\_Length} & 2 & \texttt{Protocol\_Type} & 3 & \texttt{Time\_To\_Live} & 4 & \texttt{fin\_flag\_number} \\ 
5 & \texttt{syn\_flag\_number} & 6 & \texttt{rst\_flag\_number} & 7 & \texttt{psh\_flag\_number} & 8 & \texttt{ack\_flag\_number} \\ 
9 & \texttt{ece\_flag\_number} & 10 & \texttt{cwr\_flag\_number} & 11 & \texttt{ack\_count} & 12 & \texttt{syn\_count} \\ 
13 & \texttt{fin\_count} & 14 & \texttt{rst\_count} & 15 & \texttt{HTTP} & 16 & \texttt{HTTPS} \\ 
17 & \texttt{DNS} & 18 & \texttt{Telnet} & 19 & \texttt{SMTP} & 20 & \texttt{SSH} \\
21 & \texttt{IRC} & 22 & \texttt{TCP} & 23 & \texttt{UDP} & 24 & \texttt{DHCP} \\ 
25 & \texttt{ARP} & 26 & \texttt{ICMP} & 27 & \texttt{IGMP} & 28 & \texttt{IPv} \\
29 & \texttt{LLC} & 30 & \texttt{Tot\_sum} & 31 & \texttt{Min} & 32 & \texttt{Max} \\
33 & \texttt{AVG} & 34 & \texttt{Std} & 35 & \texttt{Tot\_size} & 36 & \texttt{IAT} \\
\cline{5-8}
37 & \texttt{Number} & 38 & \texttt{Variance} & \multicolumn{4}{c|}{Total number of features = 38}  \\
\hline
\end{tabular} \vspace{-3mm}
}
\end{table*}

DoS attacks are reflected by abnormally high packet rates and repeated connection requests from a single source. It affects the services and availability of medical devices. DDoS attacks are reflected by a volume of traffic floods from multiple distributed sources, resulting in excessive bandwidth consumption. Reconnaissance attacks are reflected by short-duration probing flows and systematic port scans. MQTT attacks are reflected by unauthorized topic access and message flooding. It affects data integrity and confidentiality in the patient-monitoring systems. Spoofing attacks are reflected by identity inconsistency, such as mismatched IP–MAC pairs, unusual device IDs, and inconsistent communication patterns. Note that DoS, DDoS, Recon, and spoofing are network-based attacks, whereas MQTT is an application-based attack. The dataset comprises training and testing data for each protocol in separate files. Table~\ref{summary_feature} and Table~\ref{summary_attack} present the summary of different features and attacks present in the dataset, respectively. 

\begin{table} [t!]
\caption{Different attacks present in CICIoMT24.}
\label{summary_attack} 
\centering
{
\setlength\tabcolsep{7.0pt}
\begin{tabular}{|c|c|c|c|}
\hline
IoMT Protocol & Attack type & \multicolumn{2}{c|}{Number of samples} \\
\cline{3-4}
&  & Training & Testing \\
\hline
Bluetooth & Benign & 21750 & 6533 \\
& DoS & 99840 & 25171 \\
\cline{2-4}
& Total samples & 121590 & 31704 \\
\hline
MQTT and Wi-Fi & Benign & 19291 & 37607 \\
& DoS & 1805529 & 416676 \\
& DDoS & 4779859 & 1066764 \\
& Recon & 103726 & 27676 \\
& MQTT & 262938 & 621013 \\
& Spoofing & 16047 & 5868 \\
\cline{2-4}
& Total samples & 6987390 & 2175604 \\
\hline
\end{tabular} \vspace{-4mm}
}
\end{table}

\section{Results and Discussions} 
\label{res_dis}
This section outlines the experimental environment, evaluation methodology, and classification scenarios, followed by a detailed analysis of the experimental results for each scenario.

\subsection{Experimental Environment} 
\label{sys_set}
All algorithms are implemented in Python~3.13.6 using the Keras framework built on TensorFlow~2.2.0 and executed on a MacBook equipped with Apple M4 chip and 16~GB of RAM.

\subsection{Evaluation Methodology} 
\label{eva_method} 
To obtain reliable classification outcomes, the dataset (see Section~\ref{exp_data}) is preprocessed by handling missing values, balancing class distributions, extracting salient features, and performing binarization (only for the TM model) before training classifiers. The classifier performance is evaluated using accuracy, precision, recall, and F1-score. Accuracy measures overall correct classification. Precision indicates the reliability of detected attacks, and recall measures the IDS's ability to detect actual attacks. The F1-score balances precision and recall. Additionally, five-fold cross-validation~\cite{bhagwat2019applied} is employed to evaluate these performance metrics. The confusion matrix is used for result visualization. Further, the proposed TM-based IDS is compared with the traditional ML classifiers (see Section~\ref{mlc}) and the state-of-the-art approaches from the literature to demonstrate its effectiveness. Along this line, inference time, which is the time taken by a trained classifier to generate a prediction for a given input sample during deployment, is also evaluated and compared. To interpret the classification outcomes of the proposed TM model and enhance explainability, we present class-wise vote scores (see Eq. \eqref{eq:class_vote}) and clause activation heatmaps. Note that a higher class vote indicates stronger evidence that the input traffic belongs to class $c$, indicating a confident classification decision. The performance metrics are defined as~\cite{jaiswal2023caqoe}:
\begin{equation}
\text{Accuracy} = \frac{TP + TN}{TP + TN + FP + FN},
\end{equation}

\begin{equation}
\text{Precision} = \frac{TP}{TP + FP},
\end{equation}

\begin{equation}
\text{Recall} = \frac{TP}{TP + FN},
\end{equation}

\begin{equation}
\text{F1-score} = \frac{2 \cdot \text{Precision} \cdot \text{Recall}}
{\text{Precision} + \text{Recall}},
\end{equation}
where $TP$, $TN$, $FP$, and $FN$ denote true positive (correctly detected attacks), true negative (correctly identified benign traffic), false positive (benign traffic misclassified as attacks), and false negative (missed attacks), respectively. 

\subsection{Classification Scenarios} 
\label{class_scen}
We evaluate the classification of \texttt{benign} and \texttt{malicious attack} traffic under three different scenarios (see Table~\ref{summary_attack}). In Scenario~1,  only the Bluetooth protocol is considered, and a binary classification task is performed to distinguish between benign and attack traffic. Scenario~2 jointly analyzes MQTT and Wi-Fi protocols and performs a six-class multi-class classification to differentiate benign traffic from five distinct attack types. Additionally, in Scenario~3, all protocols are combined together to perform a seven-class multi-class classification that distinguishes benign traffic from six different attack types (five attacks from Scenario~2 and one attack from Scenario~1). In this Scenario, only the network features common to both Scenario~1 and Scenario~2 are utilized. 

\subsection{Classification Analysis of Scenario~1} 
\label{class_scen1}
\subsubsection{Data Pre-processing} 
\begin{figure}[h!]
\centering
\includegraphics[width=\columnwidth,height=4.0cm,keepaspectratio]{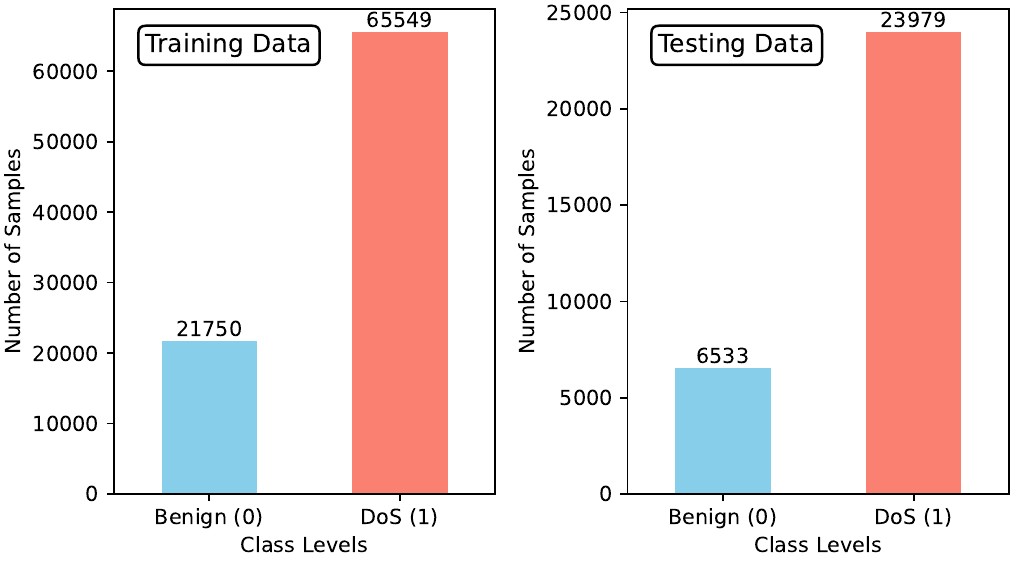} 
\caption{Class imbalance in Scenario~1: Binary classification.}
\label{clas_imb_sce1} \vspace{-5mm}
\end{figure}

\begin{figure}[h!]
\centering
\includegraphics[width=\columnwidth,height=4.0cm,keepaspectratio]{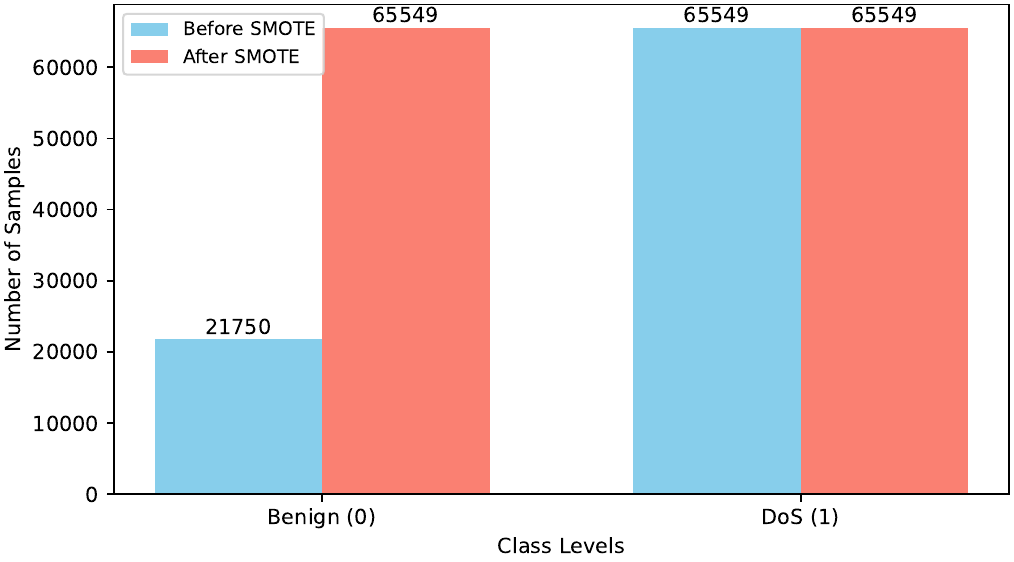} 
\caption{Balanced training class in Scenario~1.}
\label{clas_imb_sce1_smote} \vspace{-3mm}
\end{figure}

The dataset under the Bluetooth protocol (see Table~\ref{summary_attack}) exhibits class imbalance between the benign and attack (DoS) samples and may contain missing, redundant, or inconsistent information. Therefore, data cleaning is performed to ensure data quality by removing missing values and duplicate records from both the training and testing sets. After data cleaning, the dataset remains imbalanced, as illustrated in Fig.~\ref{clas_imb_sce1}. Training classifiers on imbalanced data can degrade performance and increase false positive rates. To address this issue, the Synthetic Minority Over-sampling Technique (SMOTE) is applied exclusively to the training data to generate synthetic samples for minority classes through interpolation~\cite{li2021novel}, as shown in Fig.~\ref{clas_imb_sce1_smote}. This approach reduces model bias toward the majority class and improves overall classification performance. Note that SMOTE is not applied to testing data to preserve the original data distribution and ensure unbiased model evaluation.

\subsubsection{Classifier Training} The numerical features are standardized to a common scale having zero mean and unit standard deviation. It improves training stability and enhances model performance and fairness by giving equal importance to all features. Since the TM model operates on binary inputs and learns logical rules, the standardized features are discretized into interval-based bins using the KBinsDiscretizer~\cite{bhagwat2019applied}, enabling effective binarization and interpretable clause learning. For comparison, traditional ML classifiers (see Section~\ref{mlc}) are also trained. The parameter settings and classification performance results for both the TM and ML models are presented in Table~\ref{para_scen1} and Table~\ref{res_scen1}, respectively.

\begin{table} [t!]
\caption{Parameters used for models in Scenario~1.}
\label{para_scen1} 
\centering
{
\setlength\tabcolsep{2.0pt}
\begin{tabular}{|c|c|}
\hline
Model & Parameters \\
\hline
TM & Binarizer: KBinsDiscretizer, \text{n\_bins}=5, \\ 
& encode=\text{onehot-dense}, strategy=quantile \\
\cline{2-2}
& \text{number\_of\_clauses}=100, $T$=10, $s$=2, \\
& \text{weighted\_clauses}=False, Epochs=10 \\
\hline
DT & \text{class\_weight=balanced} \\
\hline
RF & \text{class\_weight=balanced} \\
\hline
XGBoost & objective=binary:logistic, \text{eval\_metric}=logloss, \\
& \text{use\_label\_encoder}=False \\
\hline
LGBM & objective=binary, metric=\text{binary\_logloss}, \\
\hline
KNN & \text{n\_neighbors=5} \\
\hline
NB & default settings \\
\hline
LR & solver=liblinear, \text{class\_weight}=balanced \\
\hline
NN & input layer=27 neurons, first hidden layer=64 neurons, \\
& second hidden layer=32 neurons, output layer=1 neuron, \\
& hidden layer activation function=relu, \\
& output layer activation function=sigmoid, \\
& optimizer=adam, \text{loss=binary\_crossentropy}, metrics=accuracy, \\
& epochs=10, \text{batch\_size=32}, verbose=0 \\
\hline
\end{tabular} 
}
\end{table}

\begin{table} [t!]
\caption{Model performance in Scenario~1.}
\label{res_scen1} 
\centering
{
\setlength\tabcolsep{2.0pt}
\begin{tabular}{|c|c|c|c|c|c|}
\hline
Model & Accuracy & Precision & Recall & \text{F1-score} & Inference time \\
& (in \%) & (in \%) & (in \%) & (in \%) & (in microseconds \SI{}{\micro\second}) \\
\hline
TM & 0.995 & 0.999 & 0.991 & 0.995 & 0.743 \\
\hline
DT & 0.997 & 0.997 & 0.997 & 0.997 & 0.045 \\
\hline
RF & 0.998 & 0.999 & 0.996 & 0.998 & 1.930 \\
\hline
XGBoost & 0.998 & 0.999 & 0.997 & 0.998 & 0.140 \\
\hline
LGBM & 0.998 & 0.999 & 0.996 & 0.998 & 0.516 \\
\hline
KNN & 0.997 & 0.999 & 0.995 & 0.997	& 32.745 \\
\hline
NB & 0.994 & 0.997 & 0.992 & 0.994 & 0.127 \\
\hline
LR & 0.996 & 0.998 & 0.994 & 0.996 & 0.019 \\
\hline
NN & 0.997 & 0.999 & 0.994 & 0.997 & 7.673 \\
\hline
\end{tabular} \vspace{-5mm}
}
\end{table}

\begin{figure}[h!]
\centering
\includegraphics[width=\columnwidth,height=4.0cm,keepaspectratio]{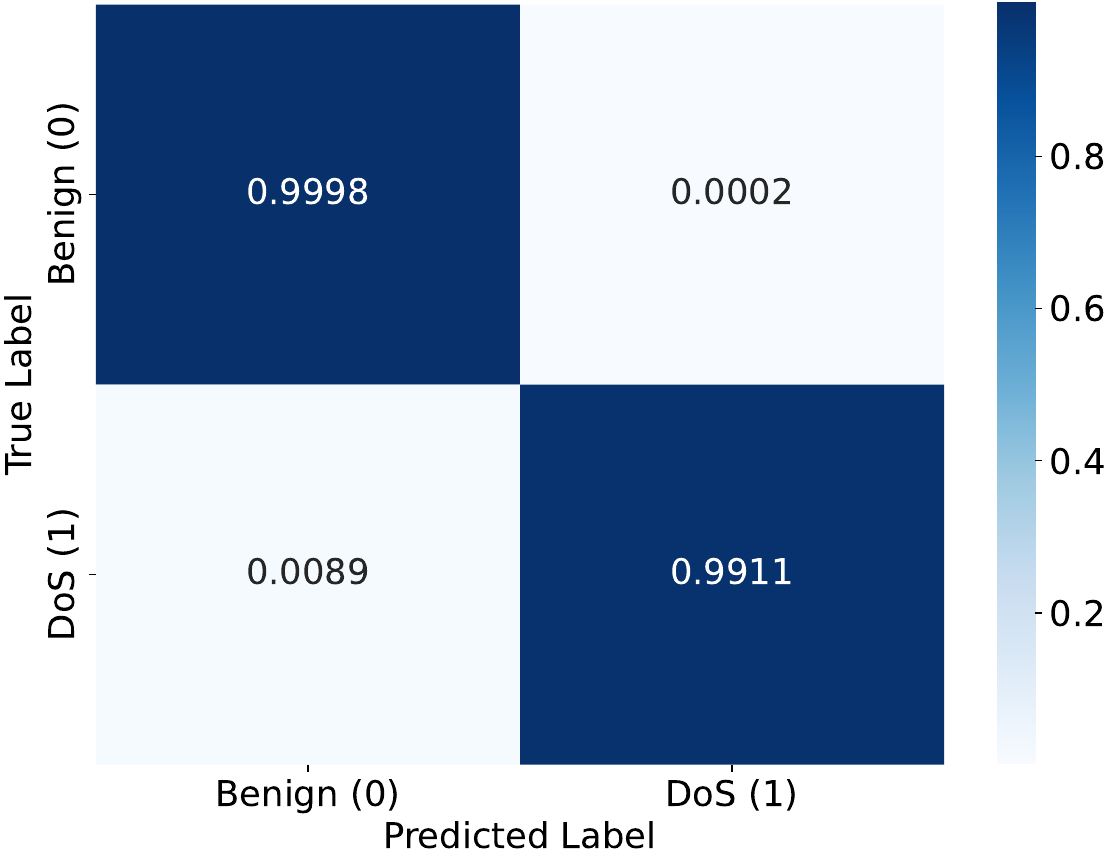} 
\caption{TM confusion matrix in Scenario~1 (two classes).}
\label{cm_TM_scen1} \vspace{-3mm}
\end{figure}

Table~\ref{res_scen1} shows that all classifier models achieve comparably high accuracy, precision, recall, and F1-score, demonstrating effective discrimination between benign traffic and DoS attacks. Logistic regression exhibits the lowest inference time of \SI{0.019}{\micro\second}. The TM model achieves an accuracy of 99.5\% with an inference time of \SI{0.743}{\micro\second}. Next, the confusion matrix of the TM model is shown in Fig.~\ref{cm_TM_scen1}. It indicates a false-positive rate of only 0.02\%, and a false-negative rate of 0.89\%, demonstrating reliable detection performance.

\subsubsection{Explainability of TM Model} 
\begin{figure}[t!]
\centering
\includegraphics[width=0.95\columnwidth,height=3.25cm,keepaspectratio]{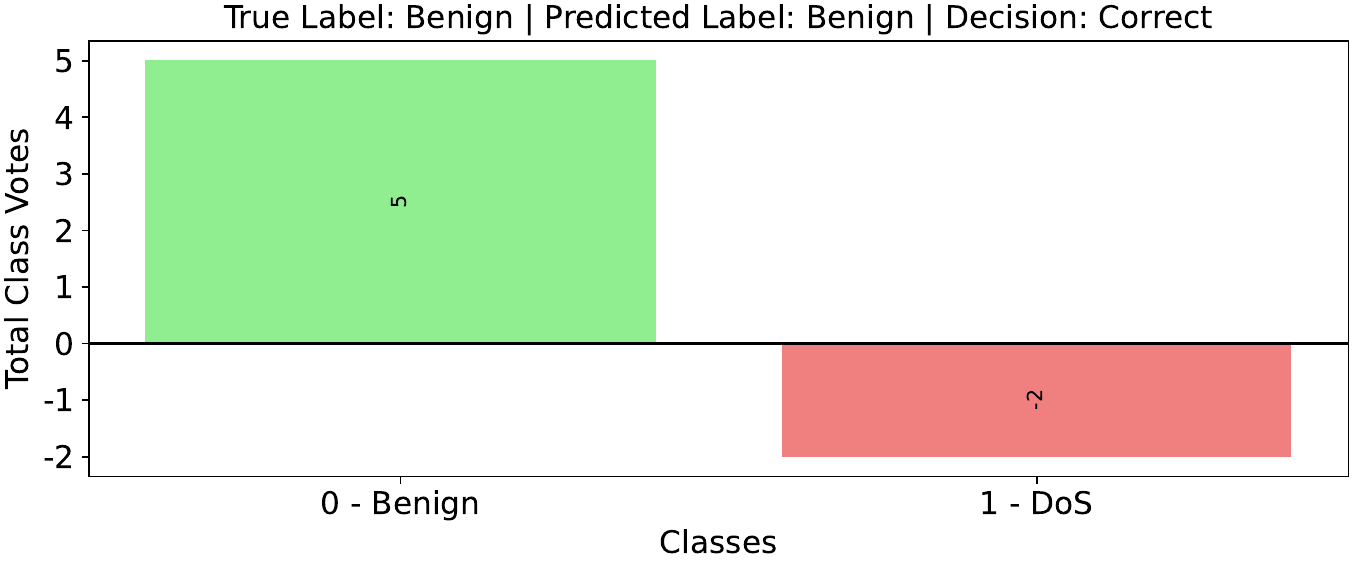} 
\caption{Class-wise votes of a Benign sample in Scenario~1.}
\label{cls_vote_scen1} \vspace{-3mm}
\end{figure}

\begin{figure}[t!]
\centering
\includegraphics[width=\columnwidth,height=3.25cm,keepaspectratio]{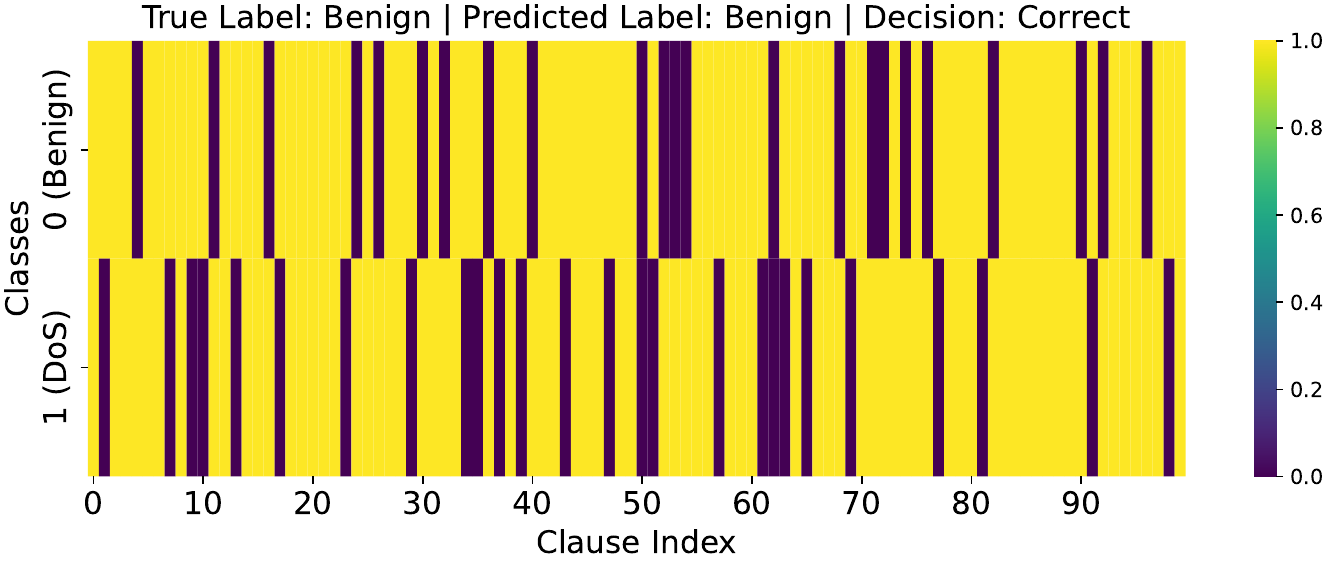} 
\caption{Clause activation heatmap of the same Benign sample.}
\label{cls_heatmap_scen1} \vspace{-5mm}
\end{figure}

To enhance the interpretability of the TM model’s decisions, Fig.~\ref{cls_vote_scen1} and Fig.~\ref{cls_heatmap_scen1} show the class-wise vote scores and clause activation heatmap for a test Benign sample, respectively. Fig.~\ref{cls_vote_scen1} shows that the Benign traffic has a higher class vote (5) than the DoS attack (-2), indicating normal traffic behaviour. In Fig.~\ref{cls_heatmap_scen1}, each cell represents the clause activation status for the given input sample, where yellow (value = 1) indicates an activated clause and dark purple (value = 0) denotes an inactive clause. For the Benign class, a larger number of positive clauses are activated, thereby contributing positively to the class vote. In contrast, the DoS class shows fewer activated clauses, resulting in a lower cumulative vote. This disparity in clause activations leads to a higher class vote for the Benign class, leading the TM model to correctly classify the input as Benign traffic. 

\begin{figure*}[b!]
\centering
\includegraphics[width=\textwidth,height=4.0cm,keepaspectratio]{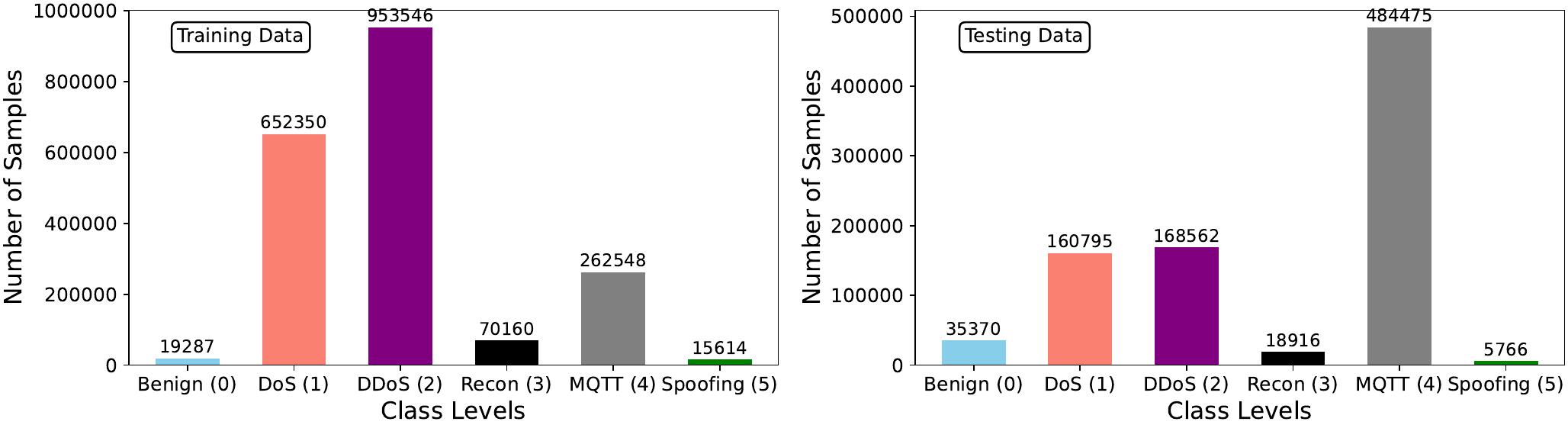} 
\caption{Class imbalance in Scenario~2: Multi-class (six-class) classification.}
\label{clas_imb_sce2} \vspace{-3mm}
\end{figure*}

\subsection{Classification Analysis of Scenario~2} 
\label{class_scen2}
\subsubsection{Data Pre-processing} 
\begin{figure}[t!]
\centering
\includegraphics[width=0.95\columnwidth,height=3.75cm,keepaspectratio]{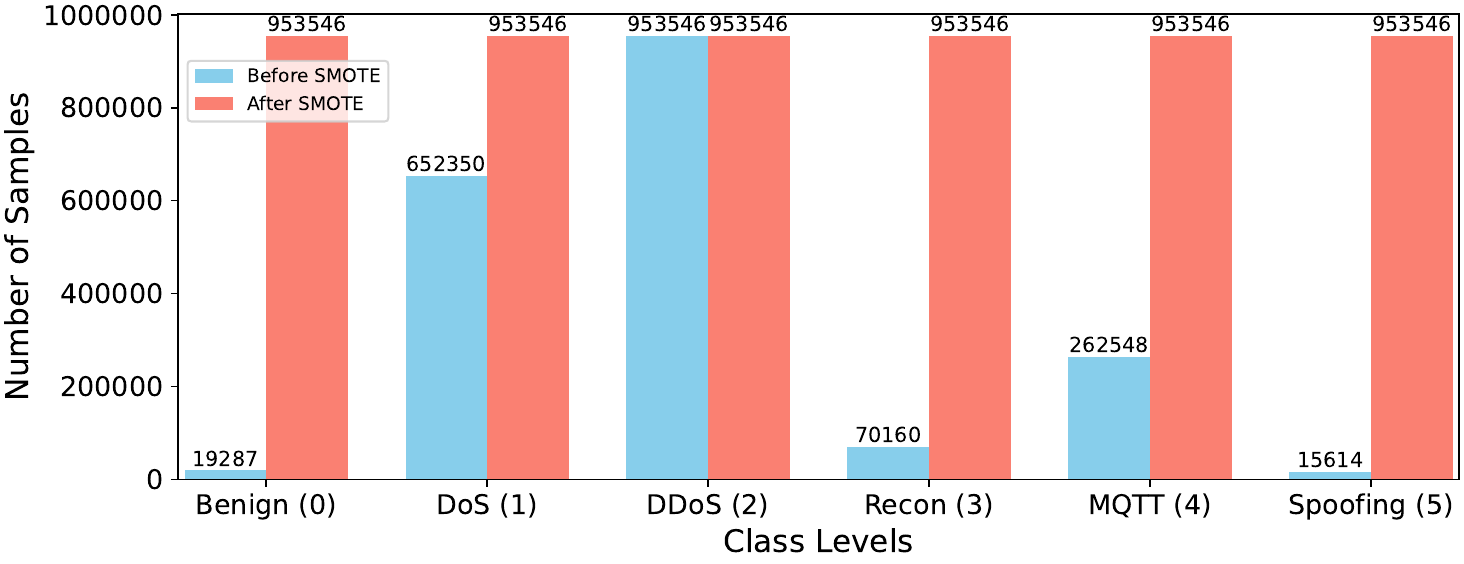} 
\caption{Balanced training class in Scenario~2.}
\label{clas_imb_sce2_smote} \vspace{-3mm}
\end{figure}

The dataset under the joint MQTT and Wi-Fi (see Table~\ref{summary_attack}) exhibits class imbalance. Therefore, it is cleaned using the same approach as in Scenario~1. The final data distribution is illustrated in Fig.~\ref{clas_imb_sce2}. For the TM model, the balanced training class data is obtained using SMOTE, as shown in Fig.~\ref{clas_imb_sce2_smote}. In contrast, for the ML classifiers, the \text{compute\_class\_weight} method~\cite{bhagwat2019applied} is employed to handle class imbalance, which assigns higher weights to the minority classes and lower weights to the majority classes in order to reduce bias toward the majority classes.

\subsubsection{Classifier Training} 
\begin{figure}[t!]
\centering
\includegraphics[width=\columnwidth,height=3.75cm,keepaspectratio]{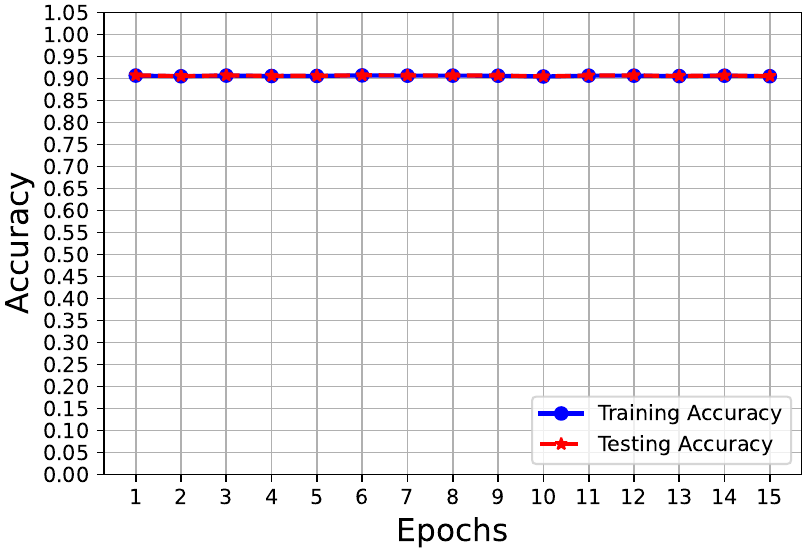} 
\caption{Training and Testing accuracy of the TM model.}
\label{train_test_acc_TM_scen2} \vspace{-2mm}
\end{figure}

The TM model is trained after data standardization and binarization, while the ML classifiers are trained after data standardization, following the same methodology as in Scenario~1. The parameters and classification performance results for both the TM and ML models are presented in Table~\ref{para_scen2} and Table~\ref{res_scen2}, respectively. Fig.~\ref{train_test_acc_TM_scen2} illustrates the training and testing accuracy across epochs, indicating proper TM model training without overfitting.

\begin{table} [t!]
\caption{Parameters used for models in Scenario~2.}
\label{para_scen2} 
\centering
{
\setlength\tabcolsep{2.0pt}
\begin{tabular}{|c|c|}
\hline
Model & Parameters \\
\hline
TM & Binarizer: The same settings as in Table~\ref{para_scen1} \\ 
\cline{2-2}
& \text{number\_of\_clauses}=100, $T$=10, $s$=5, \\
& \text{weighted\_clauses}=False, Epochs=15 \\
\hline
DT, RF & \text{compute\_class\_weight} method with \text{class\_weight=balanced} \\
\hline
XGBoost & objective=multi:softprob, \text{eval\_metric}=mlogloss, \\
& \text{tree\_method}=hist, \text{learning\_rate}=0.1, \\ & \text{max\_depth}=8, \text{n\_estimators=200} \\
\hline
LGBM & objective=multiclass, \text{learning\_rate}=0.1, \text{n\_estimators=200}, \\
& \text{compute\_class\_weight} method with \text{class\_weight=balanced} \\
\hline
KNN, NB & The same settings as in Table~\ref{para_scen1} \\
\hline
LR & solver=lbfgs, max\_iter=500, multi\_class=multinomial, \\
& \text{compute\_class\_weight} method with \text{class\_weight=balanced} \\
\hline
NN & same settings as in Table~\ref{para_scen1} except
input layer=38 neurons, \\
& output layer=6 neurons, output layer act. function=softmax, \\
& \text{loss=sparse\_categorical\_crossentropy}, \\ 
& \text{compute\_class\_weight} method with \text{class\_weight=balanced} \\
\hline
\end{tabular} \vspace{-1mm}
}
\end{table}

\begin{table} [t!]
\caption{Model performance in Scenario~2.}
\label{res_scen2} 
\centering
{
\setlength\tabcolsep{2.0pt}
\begin{tabular}{|c|c|c|c|c|c|}
\hline
Model & Accuracy & Precision & Recall & \text{F1-score} & Inference time \\
& (in \%) & (in \%) & (in \%) & (in \%) & (in microseconds \SI{}{\micro\second}) \\
\hline
TM & 0.907 & 0.913 & 0.907 & 0.906 & 4.056 \\
\hline
DT & 0.803 & 0.858 & 0.856 & 0.857 & 0.217 \\
\hline
RF & 0.816 & 0.869 & 0.875 & 0.872 & 17.588 \\
\hline
XGBoost & 0.862 & 0.868 & 0.927 & 0.890 & 1.498 \\
\hline
LGBM & 0.860 & 0.866 & 0.925 & 0.888 & 3.953 \\
\hline
KNN & 0.837 & 0.867 & 0.880 & 0.873	& 540.373 \\
\hline
NB & 0.447 & 0.596 & 0.615 & 0.474 & 0.532 \\
\hline
LR & 0.620 & 0.726 & 0.826 & 0.757 & 0.051 \\
\hline
NN & 0.743 & 0.780 & 0.876 & 0.811 & 6.443 \\
\hline
\end{tabular} \vspace{-5mm}
}
\end{table}

Table~\ref{res_scen2} demonstrates that the TM model outperforms all other classifiers, achieving over 90\% across accuracy, precision, recall, and F1-score, with an inference time of \SI{4.056}{\micro\second}. Although logistic regression yields the fastest inference time of \SI{0.051}{\micro\second}, its accuracy is limited to only 62\%. 

\begin{figure}[t!]
\centering
\includegraphics[width=\columnwidth,height=5.0cm,keepaspectratio]{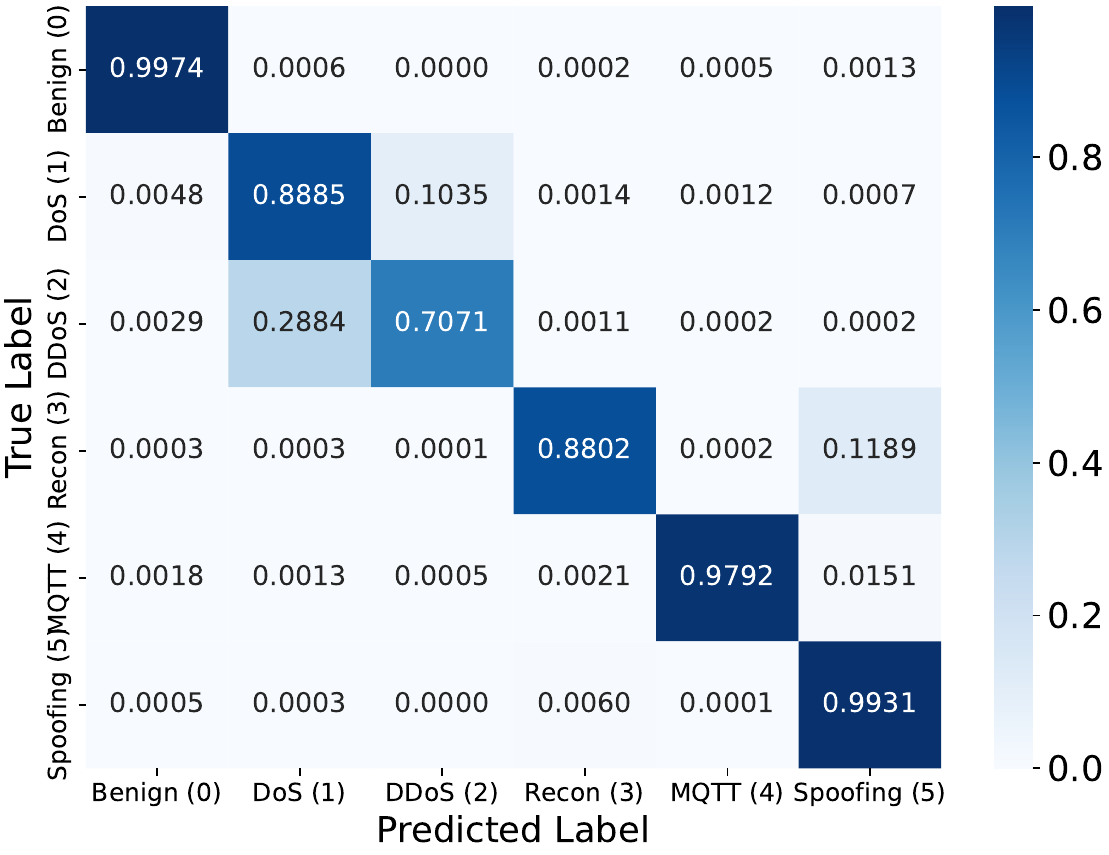} 
\caption{TM confusion matrix in Scenario~2 (six classes).}
\label{cm_TM_scen2} \vspace{-2mm}
\end{figure}

Next, the confusion matrix of the TM model is shown in Fig.~\ref{cm_TM_scen2}. It shows that the TM model achieves high classification accuracy across all six classes, with strong diagonal dominance indicating correct predictions. Benign, MQTT, and Spoofing traffic are classified with high true positive rates, while limited confusion is observed mainly between DoS and DDoS attacks due to their similar traffic characteristics.

\subsubsection{Explainability of TM Model}
\begin{figure}[t!]
\centering
\includegraphics[width=0.95\columnwidth,height=3.5cm,keepaspectratio]{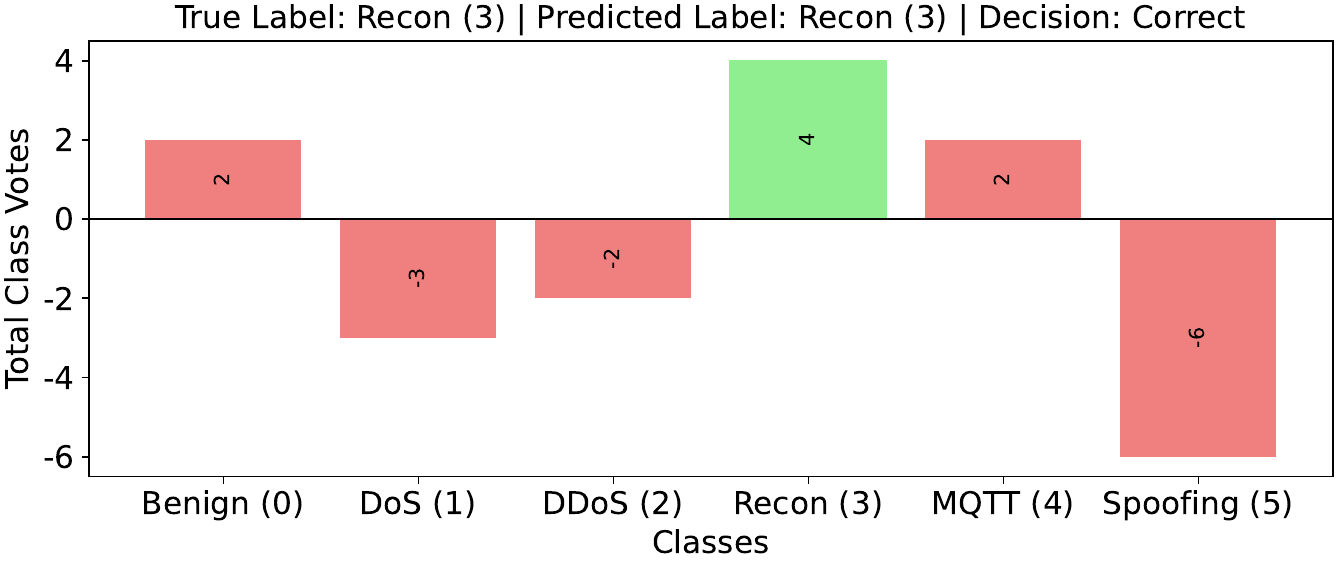} 
\caption{Class-wise votes of a Recon sample in Scenario~2.}
\label{cls_vote_scen2} \vspace{-4mm}
\end{figure}

\begin{figure}[t!]
\centering
\includegraphics[width=\columnwidth,height=3.5cm,keepaspectratio]{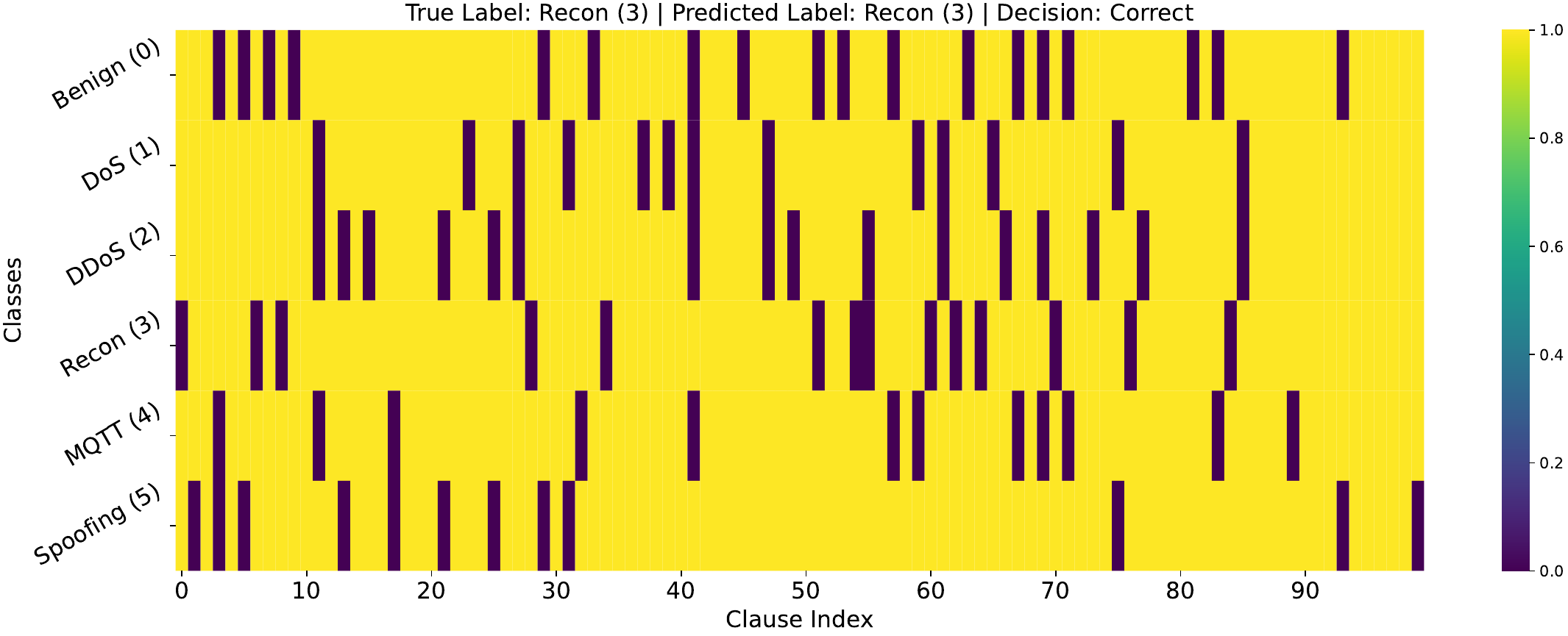} 
\caption{Clause activation heatmap of the same Recon sample.}
\label{cls_heatmap_scen2} \vspace{-2mm}
\end{figure}

Similar to Scenario~1, Fig.~\ref{cls_vote_scen2} and Fig.~\ref{cls_heatmap_scen2} show the class-wise vote scores and clause activation heatmap for a test Recon attack sample, respectively. Figure~\ref{cls_vote_scen2} shows that Recon has a higher class vote (4) than others, indicating Recon traffic behaviour. In Fig.~\ref{cls_heatmap_scen2}, a larger number of clauses are activated for the Recon class, compared to other classes. This contributes to the dominant class vote for the Recon class, leading the TM model to classify the input sample as the Recon attack correctly\footnote{Code link: https://github.com/rkj08105/TM-driven-IDS.}. 

\subsection{Classification Analysis of Scenario~3} 
\label{class_scen3}
\subsubsection{Data Pre-processing} 
\begin{figure}[t!]
\centering
\includegraphics[width=\textwidth,height=4.0cm,keepaspectratio]{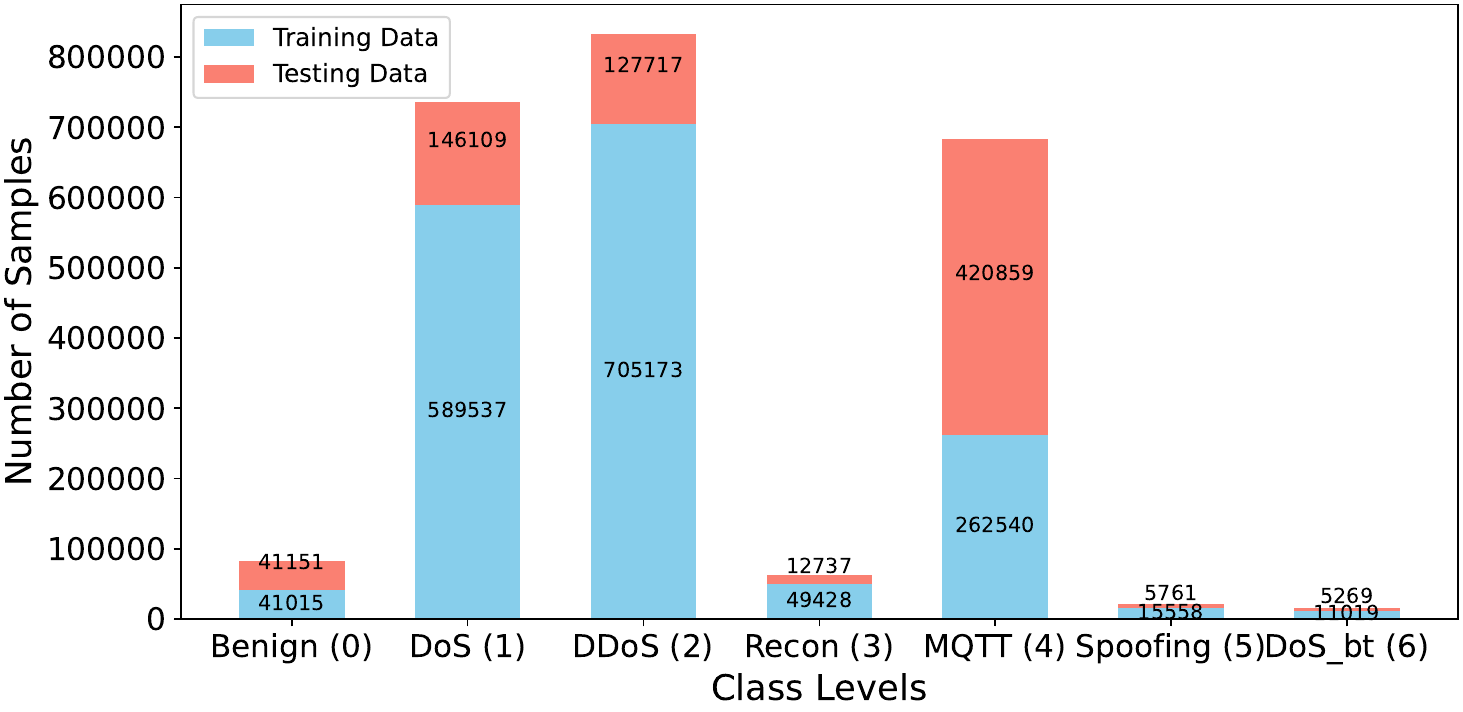} 
\caption{Class imbalance in Scenario~3: Seven classes.}
\label{clas_imb_sce3} \vspace{-2mm}
\end{figure}

The dataset combining all protocols is imbalanced and is cleaned using the same procedure as in Scenario~2, with the final class distribution shown in Fig.~\ref{clas_imb_sce3}.
SMOTE is applied to balance the training data for the TM model, while class imbalance in ML classifiers is handled using the \text{compute\_class\_weight} method~\cite{bhagwat2019applied}. Here, \texttt{DoS\_bt} refers to the DoS attack from Scenario~1 (Bluetooth).

\subsubsection{Classifier Training}  
The TM model and ML classifiers are trained using the same strategy as in Scenario~2, with parameter and performance results reported in Tables~\ref{para_scen3} and~\ref{res_scen3}, respectively. Table~\ref{res_scen3} shows that the TM model outperforms all other classifiers, achieving over 88.4\% across accuracy, precision, recall, and F1-score, with an inference time of \SI{5.413}{\micro\second}. Although logistic regression yields the fastest inference time of \SI{0.028}{\micro\second}, its accuracy is limited to 60.3\%.

\begin{table} [t!]
\caption{Parameters used for models in Scenario~3.}
\label{para_scen3} 
\centering
{
\setlength\tabcolsep{2.0pt}
\begin{tabular}{|c|c|}
\hline
Model & Parameters \\
\hline
TM & Binarizer: The same settings as in Table~\ref{para_scen2} \\ 
\cline{2-2}
& \text{number\_of\_clauses}=120, $T$=15, $s$=2, Epochs=15 \\
\hline
DT, RF, XGBoost, & The same settings as in Table~\ref{para_scen2} \\
LGBM, KNN, NB, LR &  \\
\hline
NN & The same settings as in Table~\ref{para_scen2} except \\
& input layer=11 neurons, output layer=7 neurons \\
\hline
\end{tabular} \vspace{-4mm}
}
\end{table}

\begin{table} [h!]
\caption{Model performance in Scenario~3.}
\label{res_scen3} 
\centering
{
\setlength\tabcolsep{2.0pt}
\begin{tabular}{|c|c|c|c|c|c|}
\hline
Model & Accuracy & Precision & Recall & \text{F1-score} & Inference time \\
& (in \%) & (in \%) & (in \%) & (in \%) & (in microseconds \SI{}{\micro\second}) \\
\hline
TM & 0.884 & 0.889 & 0.884 & 0.884 & 5.413 \\
\hline
DT & 0.785 & 0.863 & 0.863 & 0.863 & 0.180 \\
\hline
RF & 0.800 & 0.876 & 0.878 & 0.877 & 15.470 \\
\hline
XGBoost & 0.851 & 0.882 & 0.920 & 0.895 & 1.650 \\
\hline
LGBM & 0.788 & 0.807 & 0.888 & 0.820 & 4.319 \\
\hline
KNN & 0.813 & 0.876 & 0.877 & 0.876	& 40.287 \\
\hline
NB & 0.550 & 0.525 & 0.573 & 0.504 & 0.238 \\
\hline
LR & 0.603 & 0.689 & 0.776 & 0.722 & 0.028 \\
\hline
NN & 0.711 & 0.797 & 0.855 & 0.815 & 6.445 \\
\hline
\end{tabular} \vspace{-3mm}
}
\end{table}

\subsubsection{Explainability of TM Model}
\begin{figure}[t!]
\centering
\includegraphics[width=0.95\columnwidth,height=3.4cm,keepaspectratio]{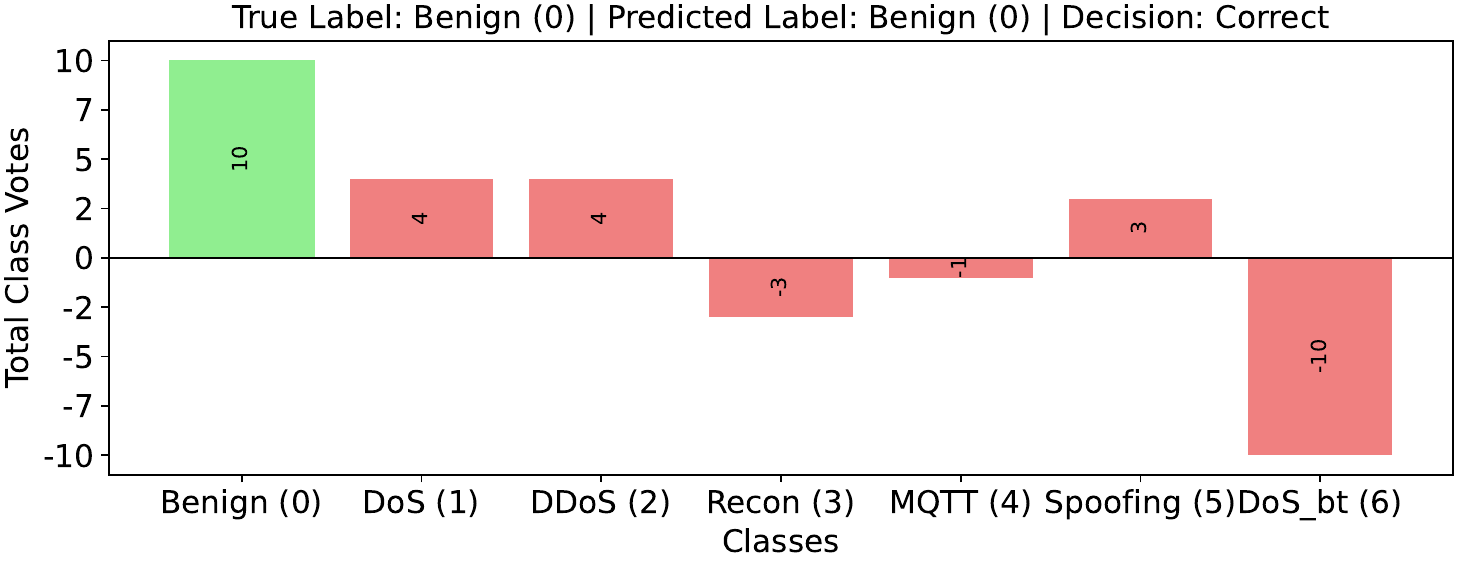} 
\caption{Class-wise votes of a Benign sample in Scenario~3.}
\label{cls_vote_scen3_sample1} \vspace{-3mm}
\end{figure}

\begin{figure}[t!]
\centering
\includegraphics[width=0.95\columnwidth,height=3.4cm,keepaspectratio]{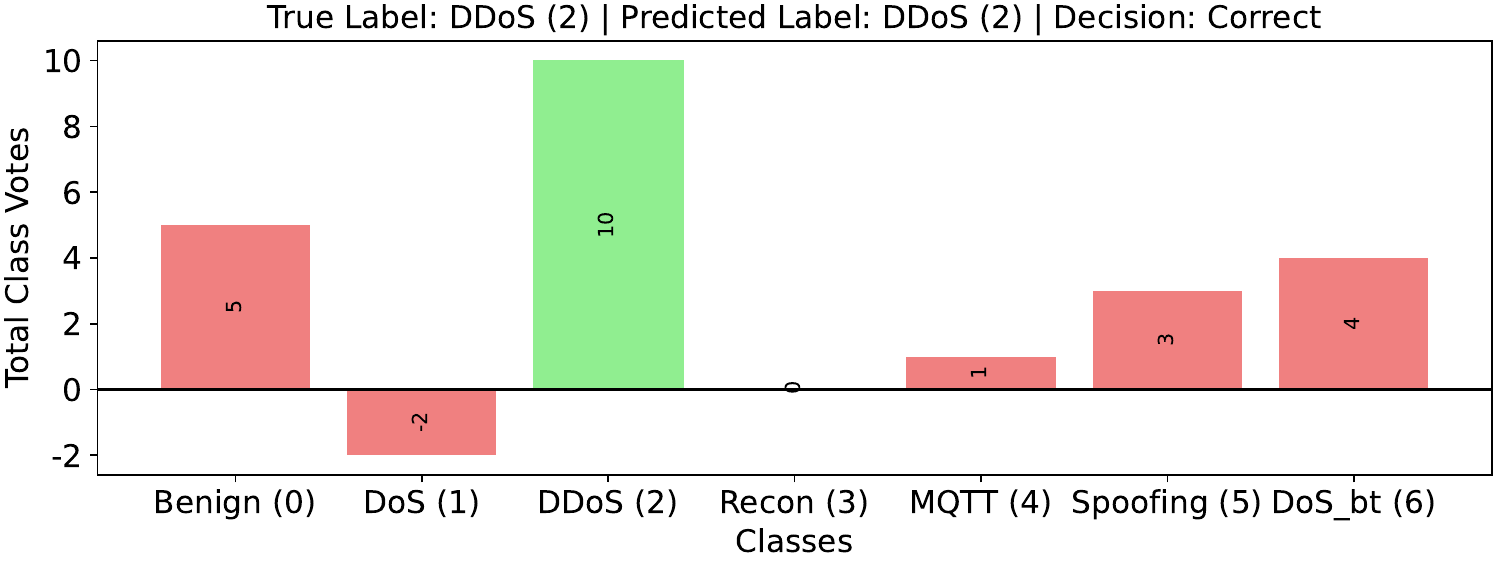} 
\caption{Class-wise votes of a DDoS sample in Scenario~3.}
\label{cls_vote_scen3_sample2} \vspace{-2mm}
\end{figure}

Fig.~\ref{cls_vote_scen3_sample1} and Fig.~\ref{cls_vote_scen3_sample2} illustrate the class-wise vote scores for Benign and DDoS attack samples, respectively, clearly indicating correct TM decisions. 

\subsection{Comparison with the State-of-the-Art Methods} 
\label{comp_sota}
Table~\ref{comp_sota_scen12} highlights the superiority of the proposed TM-based IDS in detecting benign and attacks on the same dataset.

\begin{table} [t!]
\caption{Comparison with the state-of-the-art methods.}
\label{comp_sota_scen12} 
\centering
{
\setlength\tabcolsep{3.0pt}
\begin{tabular}{|c|c|c|c|c|c|c|}
\hline
& Model & Class & Accuracy & Precision & Recall & \text{F1-score} \\
\hline
Paper~\cite{kavkas2025enhancing} & DNN/ & Binary & 0.99 & 0.99 & 0.99 & 0.99 \\
& LSTM & Six & 0.78 & 0.77 & 0.77 & 0.75 \\
\hline
Paper~\cite{dadkhah2024ciciomt2024} & ML & Binary & 0.996 & 0.971 & 0.951 & 0.961 \\
& Classifier & Six & 0.735 & 0.735 & 0.713 & 0.676 \\
\hline
Proposed & Tsetlin & Binary & 0.995 & 0.999 & 0.991 & 0.995 \\
& Machine & Six & 0.907 & 0.913 & 0.907 & 0.906 \\
\hline
\end{tabular} \vspace{-5mm}
}
\end{table}

\section{Conclusions and Future Work} 
\label{con_fut}
This paper addresses the critical cybersecurity challenge of detecting diverse cyberattacks targeting IoMT networks to safeguard patient privacy and safety. To this end, a TM-based intrusion detection system is proposed, leveraging a rule-based and interpretable ML framework that models attack patterns using propositional logic. The proposed model is trained and evaluated on the CICIoMT-2024 dataset, ensuring its effectiveness against realistic attack scenarios. Numerical results show that the proposed model performs comparably to traditional ML classifiers and state-of-the-art methods in binary classification, while surpassing them in multi-class classification, with reliability further validated through five-fold cross-validation. Moreover, the model’s decision-making process is explicitly explained, enhancing transparency and trust. Hence, our proposed IDS can provide an effective and interpretable solution for strengthening the security of IoMT networks/devices. In the future, we plan to evaluate our model on real-world network traffic collected from physical testbeds. We also intend to create a new dataset with diverse realistic attacks to develop a more robust TM-based IDS and evaluate its performance on a resource-constrained device. 

\section*{Acknowledgement}
This publication has emanated from the research project SecureIoTM: Ultra-low-energy IoT Intrusion Detection Systems using Logic-based Tsetlin Machines, under Grant Number 342167, funded by the Research Council of Norway.

\bibliographystyle{IEEEtran}
\bibliography{references}

@article{dadkhah2024ciciomt2024,
  title={{CICIoMT2024: A Benchmark Dataset for Multi-Protocol Security Assessment in IoMT}},
  author={Dadkhah, Sajjad and Neto, Euclides Carlos Pinto and Molokwu, Reginald Chukwuka and Ghorbani, Ali A},
  journal={Internet of Things},
  volume={28},
  pages={101351},
  year={2024},
  publisher={Elsevier}
}

@article{granmo2018tsetlin,
  title={{The Tsetlin Machine--A Game Theoretic Bandit Driven Approach to Optimal Pattern Recognition with Propositional Logic}},
  author={Granmo, Ole-Christoffer},
  journal={arXiv preprint arXiv:1804.01508},
  pages={1--42},
  year={2018}
}

@article{jaiswal2023caqoe,
  title={{CAQoE: A Novel No-reference Context-aware Speech Quality Prediction Metric}},
  author={Jaiswal, Rahul Kumar and Dubey, Rajesh},
  journal={ACM Trans. on Multimedia Computing, Comms. and Applications},
  volume={19},
  number={1s},
  pages={1--23},
  year={2023},
  publisher={ACM New York, NY}
}

@book{breiman2017classification,
  title={{Classification and Regression Trees}},
  author={Breiman, Leo and Friedman, Jerome and Olshen, Richard A and Stone, Charles J},
  year={2017},
  publisher={Chapman and Hall/CRC}
}

@inproceedings{jedari2015wi,
  title={{Wi-Fi based Indoor Location Positioning Employing Random Forest Classifier}},
  author={Jedari, Esrafil and Wu, Zheng and Saif, Mehrdad},
  booktitle={IEEE International Conference on Indoor Positioning and Indoor Navigation},
  pages={1--5},
  year={2015},
  organization={}
}

@inproceedings{chen2016xgboost,
  title={{Xgboost: A Scalable Tree Boosting System}},
  author={Chen, Tianqi and Guestrin, Carlos},
  booktitle={22nd ACM SIGKDD International Conference on Knowledge Discovery and Data Mining},
  pages={785--794},
  year={2016}
}

@inproceedings{ke2017lightgbm,
  title={{Lightgbm: A Highly Efficient Gradient Boosting Decision Tree}},
  author={Ke, Guolin and Meng, Qi and Finley, Thomas},
  booktitle={31st Conference on Neural Information Processing Systems},
  pages={1--9},
  year={2017}
}

@book{alpaydin2020introduction,
  title={{Introduction to Machine Learning}},
  author={Alpaydin, Ethem},
  year={2020},
  publisher={MIT press}
}

@misc{report_iomt,
  author = {},
  title = {{Internet of Medical Things Market Report (2025-2030)}},
  year = {},
  url = {https://www.grandviewresearch.com/industry-analysis/internet-of-medical-things-iomt-market-report}
}

@article{razdan2022internet,
  title={{Internet of Medical Things (IoMT): Overview, Emerging Technologies, and Case Studies}},
  author={Razdan, Sahshanu and Sharma, Sachin},
  journal={IETE Technical Review},
  volume={39},
  number={4},
  pages={775--788},
  year={2022},
  publisher={Taylor \& Francis}
}

@misc{report_threat,
  author = {},
  title = {{Global Threat Report 2025}},
  year = {},
  url = {https://www.crowdstrike.com/en-us/global-threat-report/}
}

@article{iwendi2021security,
  title={{Security of Things Intrusion Detection System for Smart Healthcare}},
  author={Iwendi, Celestine and Anajemba, Joseph Henry and Biamba, Cresantus and Ngabo, Desire},
  journal={Electronics},
  volume={10},
  number={12},
  pages={1--27},
  year={2021},
  publisher={MDPI}
}

@article{kundu360,
  title={{A 360-Degree Review of Tsetlin Machines: Concepts, Applications, Analysis, and the Future}},
  author={Kundu, Souraja and Patkar, Shruti S and Mishra, Saras Mani and Trivedi, Gaurav and Merchant, Farhad},
  journal={IEEE TechRxiv},
  pages={1--23},
  year={2025},
  publisher={IEEE}
}

@article{mitchell2014behavior,
  title={{Behavior Rule Specification-based Intrusion Detection for Safety Critical Medical Cyber Physical Systems}},
  author={Mitchell, Robert and Chen, Ray},
  journal={IEEE Trans. on Dependable \& Secure Comp.},
  volume={12},
  number={1},
  pages={16--30},
  year={2014},
  publisher={IEEE}
}

@incollection{nawaal2024signature,
  title={{Signature-based Intrusion Detection System for IoT}},
  author={Nawaal, Bakhtawar and Haider, Usman and Khan, Inam Ullah and Fayaz, Muhammad},
  booktitle={Cyber Security for Next-generation Computing Technologies},
  pages={141--158},
  year={2024},
  publisher={CRC Press}
}

@inproceedings{anitha2023artificial,
  title={{Artificial Intelligence Driven Security Model for Internet of Medical Things (IoMT)}},
  author={Anitha, Cuddapah and Komala, CR and Vivekanand, Chettiyar Vani and Lalitha, SD and Boopathi, Sampath},
  booktitle={3rd International Conference on Innovative Practices in Technology and Management},
  pages={1--7},
  year={2023},
  organization={IEEE}
}

@inproceedings{awotunde2021deep,
  title={{A Deep Learning-based Intrusion Detection Technique for a Secured IoMT System}},
  author={Awotunde, Joseph Bamidele and Abiodun, Kazeem Moses and Adeniyi, Emmanuel Abidemi and Folorunso, Sakinat Oluwabukonla and Jimoh, Rasheed Gbenga},
  booktitle={International Conference on Informatics and Intelligent Applications},
  pages={50--62},
  year={2021},
  organization={Springer}
}

@inproceedings{kavkas2025enhancing,
  title={{Enhancing IoMT Security with Deep Learning Based Approach for Medical IoT Threat Detection}},
  author={Kavkas, Nadir Can and Yildiz, Kazim},
  booktitle={IEEE International Symposium on Digital Forensics \& Security},
  pages={1--5},
  year={2025},
  organization={}
}

@inproceedings{abeyrathna2020intrusion,
  title={{Intrusion Detection with Interpretable Rules Generated using the Tsetlin Machine}},
  author={Abeyrathna, K Darshana and Pussewalage, Harsha S Gardiyawasam and Ranasinghe, Sasanka N and Oleshchuk, Vladimir A. and Granmo, Ole-Christoffer},
  booktitle={IEEE Symposium Series on Computational Intelligence},
  pages={1121--1130},
  year={2020},
  organization={}
}

@inproceedings{gunvaldsen2023towards,
  title={{Towards IoT Anomaly Detection with Tsetlin Machines}},
  author={Gunvaldsen, Ole and Thorsen, Henning Blomfeldt and Andersen, Per-Arne and Granmo, Ole-Christoffer and Goodwin, Morten},
  booktitle={IEEE International Symposium on the Tsetlin Machine },
  pages={1--8},
  year={2023},
  organization={}
}

@book{bhagwat2019applied,
  title={{Applied Deep Learning with Keras: Solve Complex Real-life Problems with the Simplicity of Keras}},
  author={Bhagwat, Ritesh and Abdolahnejad, Mahla and Moocarme, Matthew},
  year={2019},
  publisher={{Packt Publishing Ltd}}
}

@article{li2021novel,
  title={{A Novel Oversampling Technique for Class-Imbalanced Learning Based on SMOTE and Natural Neighbors}},
  author={Li, Junnan and Zhu, Qingsheng and Wu, Quanwang and Fan, Zhu},
  journal={Information Sciences},
  volume={565},
  pages={438--455},
  year={2021},
  publisher={Elsevier}
}

\end{document}